\newcommand{\CLASSINPUTtoptextmargin}{0.77in}
\newcommand{\CLASSINPUTbottomtextmargin}{0.68in}
\documentclass[conference]{IEEEtran}
\IEEEoverridecommandlockouts
\usepackage{cite}
 \usepackage[utf8]{inputenc}
     \usepackage{setspace}
\usepackage{amsmath,amssymb,amsfonts}
\usepackage{dirtytalk}
\usepackage{algorithmic}
\usepackage{graphicx}
\usepackage{balance}
\usepackage{textcomp}
\usepackage{xcolor}
\usepackage{amsfonts}
\DeclareMathOperator{\sinc}{sinc}
\usepackage{siunitx}
\usepackage{gensymb}
\usepackage[english]{babel} 
\usepackage{blindtext}
\newcommand{\RNum}[1]{\uppercase\expandafter{\romannumeral #1\relax}}

\def\BibTeX{{\rm B\kern-.05em{\sc i\kern-.025em b}\kern-.08em
    T\kern-.1667em\lower.7ex\hbox{E}\kern-.125emX}}
\begin{document}

\title{Multi-State Inter-Satellite Channel Models
}

\author{
}

\author{\IEEEauthorblockN{Eray Güven, Olfa Ben Yahia, Güneş Karabulut-Kurt} 
\IEEEauthorblockA{Poly-Grames Research Center, Department of Electrical Engineering\\ Polytechnique Montr\'eal, Montr\'eal, Canada \\
E-mail: guven.eray@polymtl.ca, olfa.ben-yahia@polymtl.ca, gunes.kurt@polymtl.ca
}
}

\maketitle

\begin{abstract}
The advancement in satellite networks (SatNets) offers vast opportunities for large-scale connectivity and flexibility. The intersection of inter-satellite communication (ISC) and the developments in rate-hungry enhanced mobile broadband (eMBB) services have resulted in potential ultra-dense deployments such as mega-constellations. The integration of inter-satellite links (ISL) with high spectrum bands requires a well-defined channel model to anticipate possible challenges and provide solutions.
In this research, we address the key concerns for defining SatNets channels and establish a comprehensive node-to-node ISL channel model for the Markov state model. The model takes into consideration satellite mobility, celestial body radiation, and cumulative thermal noise. Using this model, we analyze the bit error rate, outage, and achievable capacity of various ISC cases and propose a state-based power allocation method to improve their performance.
\end{abstract}
\begin{IEEEkeywords}
Inter-satellite links, finite-state Markov channels, channel modeling.
\end{IEEEkeywords}
\section{Introduction}
Broadband satellite networks (SatNets) will play a vital role in the next generation of wireless networks thanks to their ability to provide ubiquitous connectivity, wide coverage, and high data rates. These networks will have an ultra-dense deployment of multiple types of satellites in different orbits and with diverse capabilities. Specifically, low Earth orbit (LEO) satellites have attracted much attention as they offer improved quality of service (QoS) with lower latency and enhanced throughput compared to medium Earth orbit (MEO) and geostationary Earth orbit (GEO) due to their proximity to the Earth \cite{alsabah20216g}.

Currently, microwave bands from L-band (1-2 GHz) through Ka-band (26.5-40 GHz) and V-band (40-75 GHz) are the cornerstones for satellite communications, in addition to optical communications using lasers \cite{nie2021channel}. Although microwave bands have been efficient in data transmission for several decades, they are becoming increasingly congested as more satellites are launched into orbit. This congestion is making it difficult to maintain desired performance levels, particularly with respect to the Satellite-to-Ground Link (SGL) quality. Furthermore, the heavy interference caused by the intensively allocated spectrum to many satellites orbiting the Earth has further deteriorated the SGL quality \cite{spec}. Therefore, there is a pressing need to explore alternative technologies and techniques to improve the efficiency and reliability of satellite communications.

Inter-satellite links (ISL) liberate the SatNets from this dead-end by providing handover management, long-term connectivity, and reducing round-trip delays as a driving force. Moreover, optical/laser inter-satellite links (OISL) have become a key enabler as they can provide up to 40 Gbps throughput in free space thanks to their large and license-free spectral resource \cite{chaudhry2021laser}. As a related leading commercial development, SpaceX Starlink has started operating OISL in Antarctica recently \cite{Starlink} and the second generation satellites of OneWeb are planning to be used in OISL as well\cite{oneweb}. Nonetheless, OISLs are potentially degraded due to the high mobility, solar effects, and pointing-tracking issues \cite{tiwari}. 

In the in-depth THz band studies, the authors in \cite{thzdata} managed to reach 44 Gbps in the 300 GHz band for a line-of-sight (LoS) 1 km distance with a soft-decision forward-error correction threshold (SD-FEC) of 5.22 dB. However, it has been noted that as the distance increases, the path loss attenuation also increases to a level that becomes difficult to manage and maintain. Consequently, even though the increase in the operating frequency band remains a question mark for its use in terrestrial systems, it is expected to benefit from high-frequency communication techniques in a vacuum network (i.e. Distributed Space Systems) where the small-scale fading and minimum molecular absorption are minimal. On this basis, the lack of a realistic and flexible ISL communication channel model is one of the biggest shortcomings in studies for SatNets and this is the main source of the motivation for this study.

Previous studies on ISL have typically focused on either near Earth or deep space scenarios, but our study takes a more generalized approach by ignoring atmospheric losses such as ionospheric scintillations, Faraday rotation, rain, and fog loss. Our motivation for this research comes from the significant role ISL plays in extraterrestrial missions  \cite{harland2007space}, where loss of contact can lead to mission failure. In fact, historical data shows that 10 out of 18 satellite missions have been aborted since 1971 due to loss of contact. Hence, our study considers satellites in both Earth's orbit and extraterrestrial orbiters. Our proposed approach for analyzing ISL in formation flying scenarios offers a novel perspective, and the methods we present have the potential to improve the performance and reliability of future satellite networks.

The main contributions of this work can be summarized as follows:
\begin{itemize}
 \item We investigate the impact of relative motion among satellites during formation flying on the channel structure. We propose a model that takes into account disruptions caused by solar activities, celestial bodies, and hardware impairments such as sampling clock offset.
\item We highlight the limitations of current channel structures for ISL and propose a new method based on statistical observations. We provide a theoretical foundation for this method, which emphasizes the importance of statistical channel state information (s-CSI) for ISL networks, not only in cases where large-scale fading dominates small-scale fading but also to reduce reliance on instantaneous channel state information (i-CSI).
\item We propose a state-based ISL channel model, which allows us to analyze error bursts and outages caused by channel anomalies. We demonstrate that this model can be used to analyze different ISL scenarios in terms of bit error rate (BER), outage probability (OP), and capacity. We also discuss the key points and limitations of such networks.

\end{itemize}


The rest of the paper is organized as follows: Section \ref{sec2} presents the ISL system model including noise structure and the Sun-related effects. In Section \ref{sec3}, we define the ISL channel model and propose a stochastical model by investigating the challenges. Section \ref{sec4} and Section \ref{sec5} conclude the paper with the performance evaluation and its interpretation.

\section{System Model}
\label{sec2}
The noise characteristics of ISL links can be influenced by various factors such as the height of the nodes and the rotation of the Earth-Sun system. However, the noise effects on the ISC are not fully understood due to limitations in their applications. According to a technical report \cite{itu-r1}, the system noise temperature for an ISL system is estimated to be approximately 700 K. This value takes into account line losses, low noise, and D/C amplifiers, in addition to the noise temperature of the antenna.
\vspace{0.0cm}
\subsection {Cosmic Microwave Background (CBM) Radiation}

Studies have shown that the cosmic microwave background (CMB) radiation does not exhibit homogeneity in brightness, and its spectrum model is not steady-state due to the Sunyaev-Zeldovich effect \cite{mroczkowski2019astrophysics}. In the multipole expansion of spherical harmonics, the angular power spectrum of CMB radiation is $T(\theta)$ = $\sum_{l\zeta}{\alpha_{l \zeta} Y_{l \zeta(\theta)}}$ where $l$ number of multipole, $\zeta$ is azimuthal order, $a_{l\zeta}$ mean temperature and $Y_{l \zeta}$ fluctuations. Thus, a CMB map of different brightness can be obtained by anisotropy measurement techniques. Using the monopole terms as ($l$=0) and $Y(\theta)=1$ average CMB mean temperature of radiation becomes $T_C = 2.7255 \pm 0.0006 K$ K. For the dipole case where can be detected each spot of the sky ($l$=$1$), the amplitude drops to $3.3621 \pm0.0010$ mK. CMB rRadiation has a subtle yet persistent effect over deep space networks (DSN).


\vspace{0.0cm}
\subsection{Cumulative Noise}
\label{AA}
In addition to thermal noise, the combination of hardware noise, line loss, and impairments can also contribute to composite noise temperature. This hardware noise, which is amplified and received multiplicatively, can be expressed as
\begin{equation}
y=\sqrt{\rho}h(x+\eta)+\mathcal{W}    
\end{equation}
where $\rho$ represents the amplification coefficient, $h$ is channel coefficient for the sample of $x$, $\eta$ is the aggregated hardware distortion with $\eta\sim \mathbb{CN} (0,\kappa^2\rho )$ where $\kappa$ presents the hardware noise level \cite{bjornson2013new}, and $\mathcal{W}$ indicates the additive thermal noise with power-spectral density $N_0$. The thermal noise floor is simply defined as $\mathcal{W}_0=k_B T_{C}B$ where $k_B$ represents the Boltzmann's constant in JK$^{-1}$ and $B$ denotes the bandwidth in Hz.


\subsection{Solar Activities}
\begin{figure}[] 
    \centering
    \includegraphics[width=0.45\textwidth]{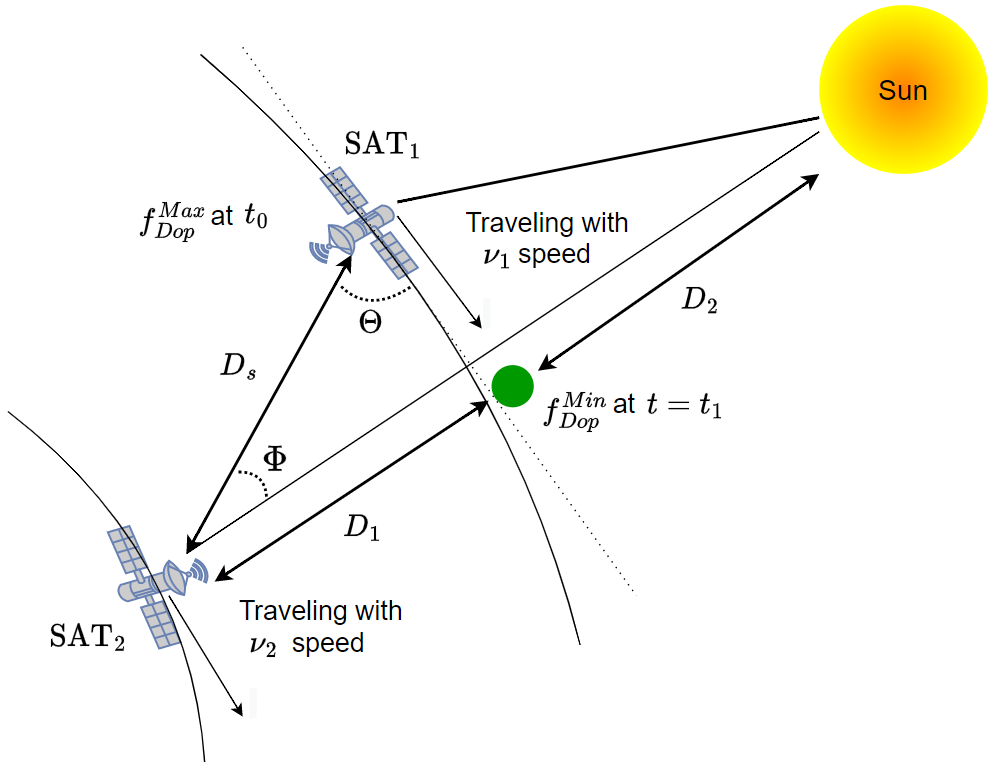}
    \caption{2-D concept of the inter-satellite communication link.}
    \vspace{-4mm}
    \label{sateclipse}
\end{figure}

The unstable solar activities are one of the main reasons why a vacuum is not just a free space environment. While solar flares follow trackable behaviors periodically, solar winds generate constantly charged electrons expressed as $\mu_p$ through space. An ISL that has the Sun in its view angle can be affected by this phenomenon heavily and has the potential to turn the LoS channel into a bursty channel. As in the experimental study of \cite{morabito2007solar}, nodes with the opposite side of the Sun (just like in solar conjunction) can cause a total void zone where the NLoS transmission case has to be taken into account. Most importantly, as highlighted in \cite{nishizuka2018deep}, sun flare prediction methods are still unable to alarm as fast as they should be.
In SatNets where size, weight, and power (SWaP) performance become more and more critical, the situation of a periodic artificial satellite eclipse occurring in the orbit has an impact on both QoS and coverage of the network. Figure \ref{sateclipse} illustrates this occurrence as the nodes are satellites that align in the same direction with the Sun with the angle of $\Theta$. This study exploits this property and takes it one step further to model an indeterminist DSN channel that correlates with $\Theta$. 

 \vspace{0.0cm}
 \section{The Challenge and the Proposed Solution}
\label{sec3}
\subsection{The Challenge}
Previous studies have revealed that ISL channel models are inadequate to account for the various effects mentioned earlier \cite{nie2021channel}. Furthermore, non-cyclostationary fluctuations in ISL channels may be induced by the radiation from celestial bodies or hardware, causing sudden drops in the signal-to-noise ratio (SNR) \cite{ho2008solar}.

\begin{figure}[t] 
    \centering
    \includegraphics[width=0.45\textwidth]{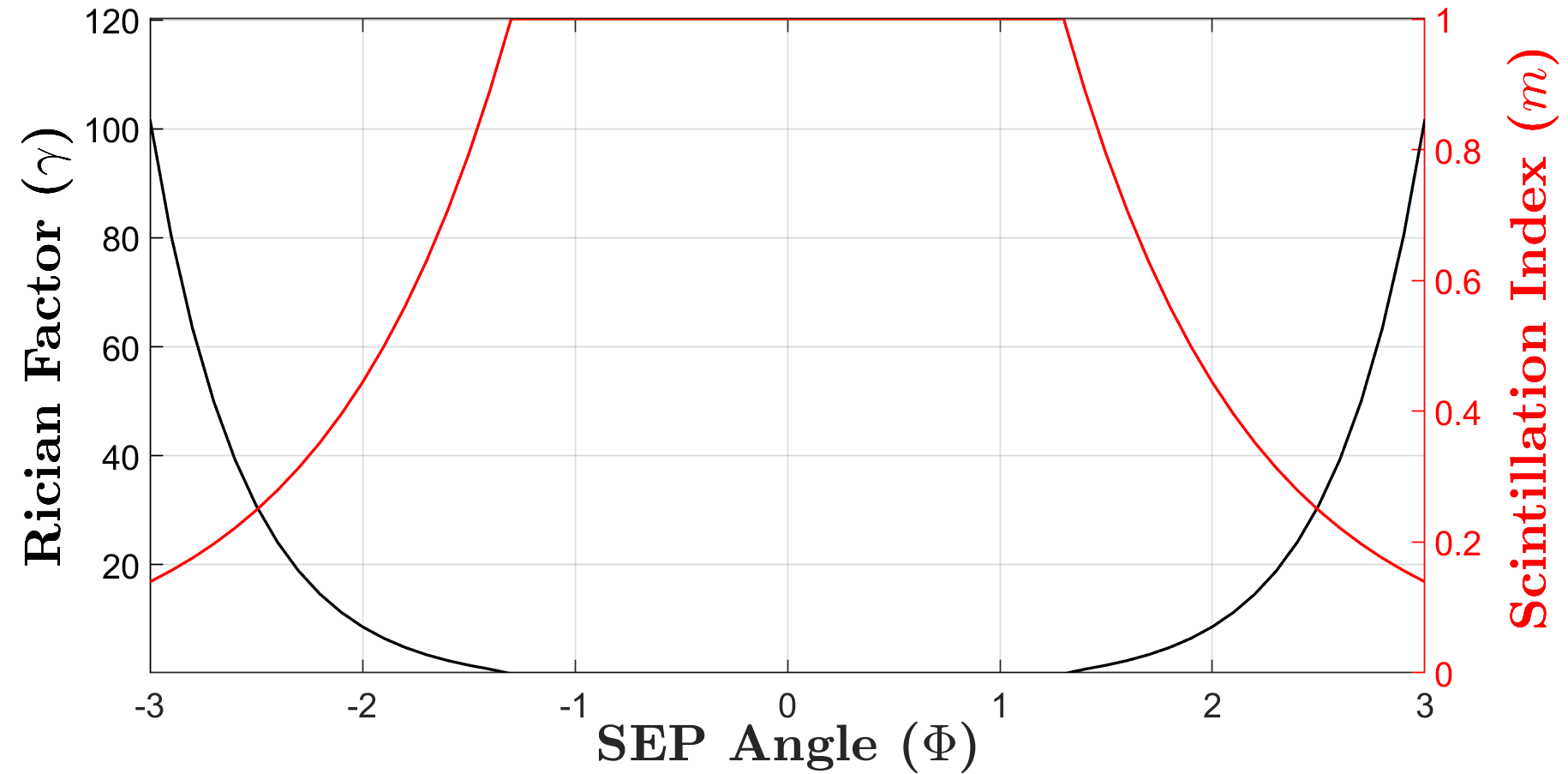}
    \caption{The relationship of the scintillation index with the channel with respect to its angular position.}
    \vspace{-4mm}
    \label{SEP}
\end{figure}

\begin{figure}[b] 
    \centering
    \vspace{-4mm}
    \includegraphics[width=0.45\textwidth]{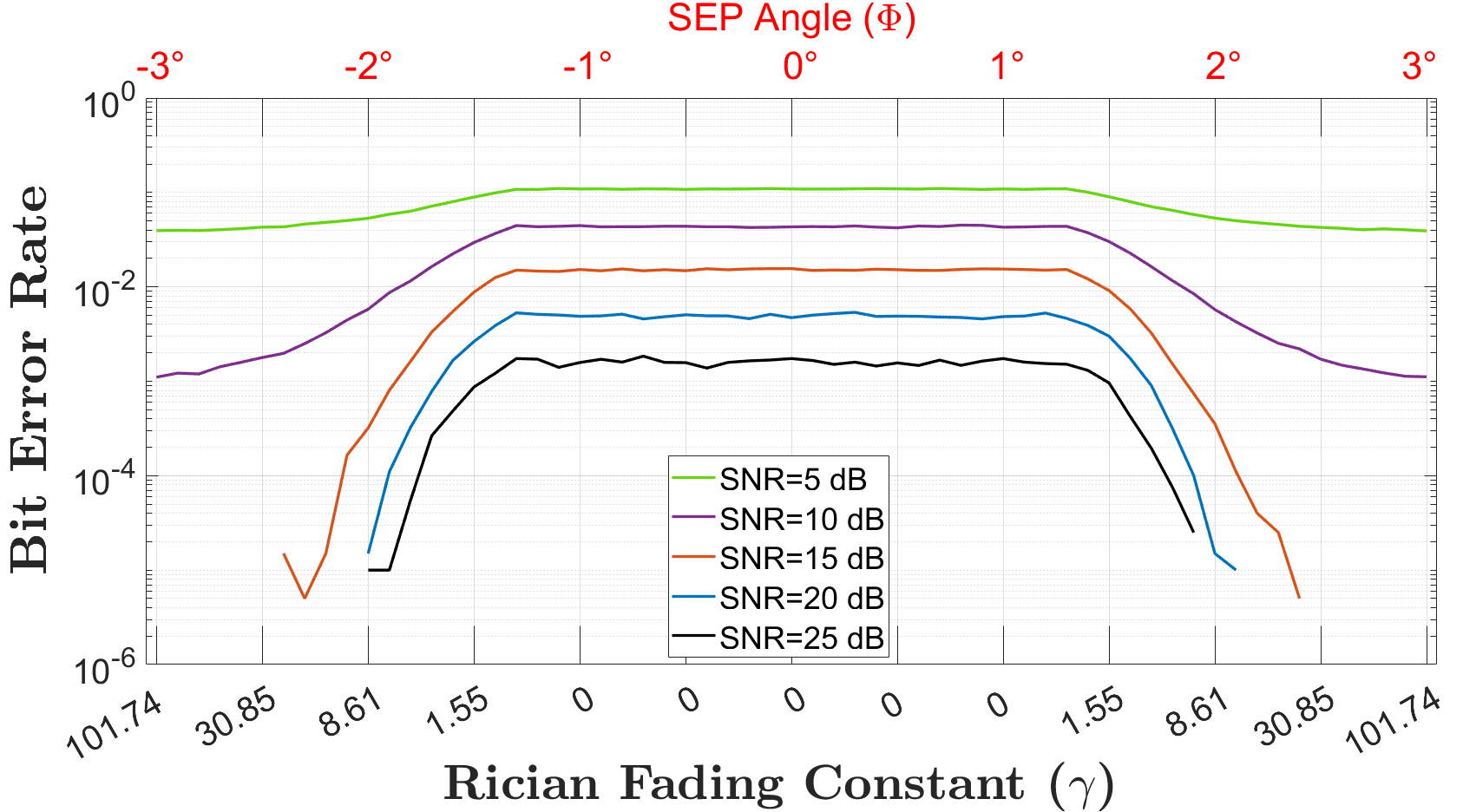}
    \caption{BER with respect to the SEP angle of the ISL. }
    \vspace{-4mm}
    \label{berangle}
\end{figure}

This study aims to highlight the limitations of using a purely deterministic or spatial channel model, as it would not be realistic, applicable, or flexible enough. It has been shown that the probability distribution of the scintillation magnitudes, resulting from the solar flare effect where a DSN ISC has Sun-Earth Probe (SEP) angle $\Phi$ with the Sun, is correlated with the Rice distribution \cite{feria1997solar}. Thus, the effect of the Sun on the ISL channels under a more realistic scenario can be simply modeled for free space. The considered effects in this model are free space loss (PL) and solar scintillation, whose amplitude is defined by the Rician distribution with $p_x(x|v,\sigma)$ and associated with the Rician fading constant ($K-$Factor) $\gamma={v^2}/{2\sigma^2}$. In addition, the effect of free space attenuation can be expressed as
\begin{equation}
\label{Eq3}
    \mathrm{PL}(dB)=32.45 + 20\log_{10}{f} + 20\log_{10}{d},
\end{equation}
where $f$ is the carrier frequency (MHz) and $d$ indicates the distance (km) with the interval $[D_1, D_s]$.
As discussed before, the relative signal intensity fluctuations can be characterized by the scintillation index, $m$, which is the root mean square of the signal intensity fluctuation divided by the mean signal intensity. This parameter characterizes the magnitude of electron density fluctuations in weak scintillation smaller than Fresnel zone size, $\sqrt{\lambda z}$ where $z$ is the effective scattering distance of the solar plasma charged $\mu_p$ \cite{morabito20001998}. Depending on the operation band, the scintillation impact caused by the $\Phi$ can also change. As a result, channel fluctuations for $-\xi<\Phi<\xi$ can be investigated in the interested band. 
Note that $m$ $\in$ $\mathbb{R}$ $\rightarrow$ $[0,1]$ and $m=0$ is the weakest scintillation case whereas $m=1$ is the strongest scintillation case. The relationship between $\gamma$ and $m$ can be written as in \cite{pan2018review}
\begin{equation}
\label{Eq4}
    \gamma=\frac{\sqrt{1-m^2}}{1-\sqrt{1-m^2}}.
\end{equation}
The scintillation index is related to the satellite's position relative to the Sun. This relationship can be expressed as follows
\begin{equation}
\label{Eq5}
    m=
    \begin{cases}
        e^{-a_1(\Phi-\theta_0)+a_2(\Phi-\theta_0)}, & \Phi<\theta_0 \\
        1, & \Phi>\theta_0 ,
    \end{cases}
\end{equation}
where $a_1= 1.14 \pm 0.09$, $a_2= 0.02 \pm 0.02$, and $\theta_0 \approx -1.3$ degree. Therefore it is possible to relate the s-CSI to $m$ and thereby to the SEP angle $\Phi$. Figure \ref{SEP} illustrates the variation of $m$ with respect to $\Phi$. The impact of $m$ can be directly observed by examining the angle-varying BER performance, as shown in Figure \ref{berangle}.

It is worth noting that in the coverage area (as shown in Figure \ref{sateclipse}), the LoS between Sat$_1$ and Sat$_2$ remains unobstructed at all times. The initial distance between Sat$_1$ and Sat$_2$ is denoted by $D_s$, and the relative speed of Sat$_2$ with respect to Sat$_1$ is denoted by $\nu_{21}$, while the relative speed of Sat$_1$ with respect to Sat$_2$ is denoted by $\nu_{12}$ (where $|\nu_{21}|=|\nu_{12}|$). It should be noted that any corruptive effects, such as the Doppler shift, can be modeled over a flat fading channel. To account for the realistic effects of the ISL channel, the Doppler shift of the LoS component in each path is examined using Jakes' Doppler spectrum \cite{jeruchim2006simulation}.

\subsubsection{Doppler frequency} 
It is important to note that at the moment when the satellites enter the coverage area, the Doppler frequency reaches its maximum value, denoted as $f_{Dop}^{Max}$, due to the maximum variation of $\nu_{21}$ in this region. On the other hand, the minimum Doppler frequency ($f_{Dop}^{Min}$) is observed at the closest distance $D_1$ from the satellites, as shown in Figure \ref{sateclipse}. It should be noted that the value of $f_{Dop}$ depends on the elevation angle $\theta$, which is considered to be constant and equal to zero for the purposes of this study.


From a relativistic point of view from Sat$_2$ as a stationary ($\nu_2 = 0$) and transmitter Sat$_1$ emmiting the wave with $f_{S1}$ along its path during $[t_0,t_2]$ \cite{pan2018review}
\begin{equation}
\label{Eq6}
 \begin{cases} 
       f_{S2}(t)=f_{S1}(c/(c-\nu_{S1}(t))) & t_0 \leq t < t_1 \\
      f_{S2}(t)=f_{S1}(c/(c+\nu_{S1}(t))) & t_1 \leq t \leq t_2, 
   \end{cases}
\end{equation}
where $f_{S2}$ indicates the observed frequency at Sat$_2$ and $c$ is the speed of light. Note that $t_{0,2}$ is associated with $f_{Dop}^{Max}$ whereas $t_1$ corresponds to $f_{Dop}^{Min}$. As a result, $f_{Dop}^{Max}=(\nu_{t_{0,1}}f_{S1})/c$ creates the Jakes' power spectrum density as follows 
\begin{equation}
    S(f_{S2})=\frac{1}{\sqrt{1-(f_{S2}/f_{Dop}^{Max})^2}}, |f_{S2}|<|f_{Dop}^{Max}|.
\end{equation}
Using the obtained $h_k$ coefficients from $S(f)$, a time-varying Doppler shift can be obtained by using a finite impulse response filter for each different $\Phi$.
It is obvious that the angular approach and recession result in a transverse Doppler shift \cite{morin2008introduction}. As a consequence, it is possible to rederive $f_{S2}$ by $\Theta$, inherently SEP angle $\Phi$. The travel time of the wave from Sat$_2$ to Sat$_1$ is denoted as $\overline{t}$, and $\Theta$ can be obtained as \cite{pan2018review}
\begin{equation}
\label{Eq7}
    \cos({\Theta}) = \frac{\nu_{S1}(t_1 - t_0)}{c \overline{t}}.
\end{equation}
As a rule of thumb, by using $\Theta = (\pi/2)-\Phi$ for $[t_0,t_1]$, (\ref{Eq6}) can be therefore expressed as
\begin{equation}
\label{Eq8}
 \begin{cases} 
    f_{S2}(t)=f_{S1}(c/(c-(\nu^2_{S1}(t)\cos((\pi/2)-\Phi)))) & t_0 \leq t < t_1, \\
    
    f_{S2}(t)=f_{S1}(c/(c+(\nu^2_{S1}(t)\cos((\pi/2)-\Phi)))) & t_1 \leq t  \leq t_2.
    \end{cases}
\end{equation}
\subsubsection{Channel model}
In addition, the end-to-end link is single-input, single-output (SISO), and the Rician fading channel is structured as a linear finite impulse response filter. Therefore, the output of channel $h_i$ can be given as \cite[Sect. (3)]{jeruchim2006simulation}
\begin{equation}
    \label{eqout}
    y_i=\sum_{n=-N_1}^{N_2}  s_{(i-n)} h_n \sinc(\frac{\tau_p}{T_s}-n),
\end{equation}
where $h_n \in$ \textbf{h} $= \{{h_1}, {h_2},...,{h_D}\}^T$, $T_s$ is the input sample period, $s_i$ is the $i$-th sample data, $\tau_p$ is the path delay, and $N_1$, $N_2$ denote the time intervals. In-phase and quadrature components are used in the process of creating a Rician fading envelope with the Sum-of-Sinusoid technique as $z(t)=z^{(1)}(t) + jz^{(2)}(t)$,
where $j=\sqrt{-1}$ and $ z^{(i)}(t)$ waveform can be shown as \cite[Sect. (3)]{jeruchim2006simulation}
\begin{equation} 
    z^{(i)}(t) = \sqrt{\frac{2}{N}}\sum_{n=1}^N \cos{(2\pi f_{Dop,n}^{(i)}t\cos(\overline{\theta}^{(i)}_n)+\psi_n^{(i)})},  i=1,2.
\end{equation}
$N$, $\overline{\theta}$ and $\psi$ denote the number of sinusoids in a single path, the phase of the corresponding component in $z^{(i)}(t)$ and the phase which is an i.i.d. random variable with uniform distribution over the interval (0,2$\pi$] for each path respectively.
Thus, the average ISL path gain with the channel impulse response (CIR) coefficients can be extracted by using $z^{(i)}(t)$. While $\Omega = E[|\textbf{h}|^2]$ is the channel gain, the fading process can be defined as $\textbf{h}=\sqrt{\Omega}z$. It can be concluded for each path along with $\Phi \in [-\xi, \xi]$ as following \cite[Sect. (3)]{jeruchim2006simulation}
\begin{equation}
    \textbf{h}_{\Phi}=\sqrt{\Omega_{\Phi}} \biggl( \frac{z_{\Phi}}{\sqrt{\gamma_{\Phi} + 1}}+ \sqrt{\frac{\gamma_{\Phi}}{\gamma_{\Phi} + 1}}e^{j(2\pi f_{Dop,{\Phi}} t)} \biggl).
\end{equation}
Therefore, $y_i$ for each $\textbf{h}$ can be extracted by Eq.(\ref{eqout}) in continuation. As a consequence, the random variable for Rician fading samples can be shown as $h=X+jY$ where $X, Y\sim$ $\mathbb{N}(\mu,\sigma^2), z^{(1)}\subset X, z^{(2)}\subset Y$.
\subsection{Proposed $K$-state ISL Channel Model}
Due to the issue of non-stationary channel conditions in ISL links, we propose a flexible approach to handle such dynamic conditions that cannot be predicted or interpreted statistically. Our approach involves categorizing the regional or temporal effects of different signal states and combining them into a more comprehensive and unified process through Markov states.
A memoryless and homogeneous finite set of states can be defined as $S = \{s_0, s_1,..., s_{K-1}\}$ with $\{S_n\},n=\{0,1,2,..., K-1\}$ are the constants for Markov process. Hereunder, the transition probability property can be written as
\begin{equation}
    t_{j,k}=Pr(S_{n+1}=s_k|S_n=s_j), n=\{0,1,2,..., K-1\},
\end{equation}
where $j,k \in \{0,1,2,..., K-1\}$. To conclude, using $t_{j,k}$ a Markov transition probability process is denoted as $\bold{T}^{K\times K}$. 
The probability of $k$th-state for any time can be written as follows
\begin{equation}
\begin{split}
p(k)&=\mathbb{P}(S_n=s_k), k\in \{0,1,2,..., K-1\}, \\
    &=\sum^{K-1}_{j=0}t_{j,k}p_j
\end{split}
\end{equation}
The validity of the probability axioms in the equivalence of each channel to a $\textbf{T}^{K\times K}$ can be given as
\begin{equation}
\begin{split}
&0<p(k)\leq 1, \forall k \in \{0,1,2,..., K-1\}, \\
&\sum_{k\in S}p(k)=1.
\end{split}
\end{equation}
In conclusion, the channel structure modeled with $f_{Dop}$ and $m$ can be separated into $K$-states with different statistical properties, hence Sat$_1$ and Sat$_2$ can adapt to the behavior of each state with different approaches.

 \begin{table}[]
   \centering
\caption{Input Parameters} 
  \resizebox{0.9\linewidth}{!}{
\label{tab1}
\begin{tabular}{|lclll|}
\hline
\multicolumn{5}{|c|}{\textbf{Parameters}}                                                                                                                \cr \hline \hline
\multicolumn{1}{|l|}{\textbf{Common Parameters}}      & \multicolumn{4}{c|}{\textbf{Values}}                                                                      \cr \hline
\multicolumn{1}{|l|}{Modulation}             & \multicolumn{4}{c|}{4QAM}                                                                  \cr \hline
\multicolumn{1}{|l|}{Number of Samples ($N_s$)}                & \multicolumn{4}{c|}{$10^6$}                                                                     \cr \hline
\multicolumn{1}{|l|}{Bandwidth ($B$) [Hz]}                   & \multicolumn{4}{c|}{$10^6$}                                                                        \cr \hline
\multicolumn{1}{|l|}{Frequency ($f_{S1}$) [GHz]}                   & \multicolumn{4}{c|}{10}                                                                        \cr \hline
\multicolumn{1}{|l|}{$T_C$ [\degree K]}                     & \multicolumn{4}{c|}{$2.7255 \pm 0.0006$}                                                                          \cr \hline
\multicolumn{1}{|l|}{Sample Time ($T_s$)}                     & \multicolumn{4}{c|}{$10^{-5}$}                                                                          \cr \hline
\multicolumn{1}{|l|}{Path Loss Exponent ($\eta$)}                   & \multicolumn{4}{c|}{2}                                                                        \cr \hline
\multicolumn{1}{|l|}{$\kappa$}                   & \multicolumn{4}{c|}{0.05}                                                                        \cr \hline
\multicolumn{1}{|l|}{Elavation ($\theta$) [\degree]}              & \multicolumn{4}{c|}{0}                                                                            \cr \hline
\multicolumn{1}{|l|}{s-CSEE ($\sigma^2_{sCSEE}$)} & \multicolumn{4}{c|}{0}                                                                            \cr \hline \hline
\multicolumn{1}{|l|}{\textbf{Angle Varying}}  & \multicolumn{4}{c|}{ }                                                                            \cr \hline \hline
\multicolumn{1}{|l|}{Azimuth Interval ($\Phi$) [\degree]}              & \multicolumn{4}{c|}{$-3\degree< \Phi <3\degree$}                                                                   \cr \hline
\multicolumn{1}{|l|}{SNR Interval [dB]}           & \multicolumn{4}{c|}{[0, 30]}                       \cr \hline
\multicolumn{1}{|l|}{Number of States ($K$)}           & \multicolumn{4}{c|}{1}                                                                \cr \hline
\multicolumn{1}{|l|}{$D_s$ [km]}           & \multicolumn{4}{c|}{10}                                                                \cr \hline
\multicolumn{1}{|l|}{$\nu_{12}$ [km/s]}               & \multicolumn{4}{c|}{0.1}                                                                    \cr \hline \hline
\multicolumn{1}{|l|}{\textbf{Time  Varying}}          & \multicolumn{1}{c|}{\textbf{Case \RNum{1}}} & \multicolumn{1}{c|}{\textbf{Case \RNum{2}}} & \multicolumn{1}{c|}{\textbf{Case \RNum{3}}} & \multicolumn{1}{c|}{\textbf{Case \RNum{4}}}  \cr \hline \hline
\multicolumn{1}{|l|}{$\nu_{12}$ [km/s]}              & \multicolumn{1}{l|}{2}       & \multicolumn{1}{l|}{2}       &
\multicolumn{1}{l|}{2}       & \multicolumn{1}{l|}{4}       \cr         \hline
\multicolumn{1}{|l|}{Number of States ($K$) }              & \multicolumn{1}{l|}{3}       & \multicolumn{1}{l|}{3}       & \multicolumn{1}{l|}{3}       & \multicolumn{1}{l|}{3}       \cr         \hline
\multicolumn{1}{|l|}{$\tau_T$ [s]}              & \multicolumn{1}{l|}{10}       & \multicolumn{1}{l|}{10}       &
\multicolumn{1}{l|}{10}       &
\multicolumn{1}{l|}{10}       \cr         \hline
\multicolumn{1}{|l|}{Azimuth ($\Phi$) [$\degree$]}              & \multicolumn{1}{l|}{30}       & \multicolumn{1}{l|}{2}       &
\multicolumn{1}{l|}{0}       &
\multicolumn{1}{l|}{30}       \cr         \hline
\multicolumn{1}{|l|}{$D_s$ [km]}               & \multicolumn{1}{l|}{69.28}       & \multicolumn{1}{l|}{60.03}       &
\multicolumn{1}{l|}{60}       &
\multicolumn{1}{l|}{69.28}      \cr         \hline
\multicolumn{1}{|l|}{$\gamma$}         & \multicolumn{1}{l|}{$\infty$}       & \multicolumn{1}{l|}{8.6193}       &
\multicolumn{1}{l|}{0}       &
\multicolumn{1}{l|}{$\infty$}       \cr        \hline
\multicolumn{1}{|l|}{Initial State ($p(k)_{t=0}$)}                 & \multicolumn{1}{l|}{2}       & \multicolumn{1}{l|}{2}       &
\multicolumn{1}{l|}{2}       &
\multicolumn{1}{l|}{2}       \cr         \hline
\multicolumn{1}{|l|}{$\Gamma_k$ [bps/Hz]}     & \multicolumn{1}{l|}{2}       & \multicolumn{1}{l|}{2}       &
\multicolumn{1}{l|}{2}       &
\multicolumn{1}{l|}{2}      
\cr
\hline 
\end{tabular}}
\vspace{-4mm}
\end{table}

\subsection{Example: Three-state ISL Model }
In order to describe an ISL channel state with Markov chains, the $K=3$ states channel model with $k \in S_2 = \{0,1,2\}$ has been investigated in this study. This model defines three states as \say{GOOD}, \say{MODERATE}, and \say{BAD} where the boundaries of separation and characterization of the states are determined with respect to the $\Gamma_\Omega$. More in details, the state with the perfect channel moment at the time of communication link is \say{GOOD}, or \say{BAD} when there is interference in the signal due to one of the external factors mentioned before. The state where the channel is at swing but the link is still available for data transfer is defined as the \say{MODERATE} state. The transition probabilities are labeled as $p_{ab} \in \mathbb{R}$ and $a,b = \{1,2,3\}$.

The three-state Markov model for ISL networks was chosen in this study since it is more realistic than the 2-state Gilbert Eliott model \cite{7490954} and has reasonably low complexity to be applied practically. 


The channel structures for three-state Markov states are given as follows 
\begin{multline}
\textbf{h}_{\Phi,k}=\sqrt{\Omega_{{\Phi,k}}}\biggl(\frac{z}{\sqrt{\gamma_{\Phi,k} + 1}}+\\ \sqrt{\frac{\gamma_{\Phi,k}}{\gamma_{\Phi,k} + 1}}e^{j(2\pi f_{Dop_{\Phi,k}} t)}\biggl) , k\in S.
\end{multline}

\begin{figure}[b] 
    \centering
    \includegraphics[width=0.45\textwidth]{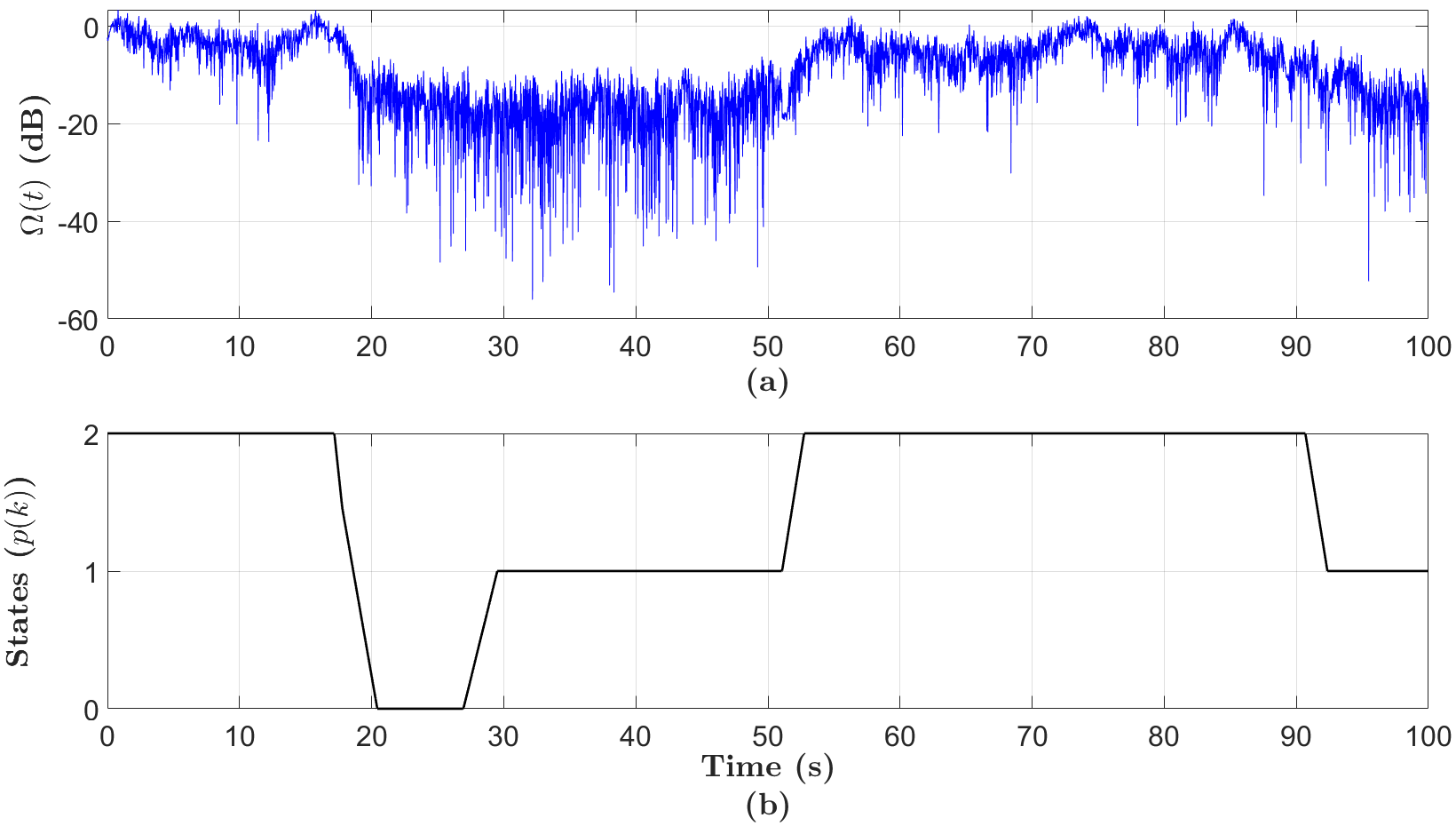}
    \caption{The state series for $\tau_T=100 s$ corresponding to channel gain.}
    \vspace{-4mm}
    \label{100s}
\end{figure}

The probability of the system being in any of the three states is certain as $Pr(f_i)=1$ and the CIR of the $p(k),\forall k$ are divided by the $\Gamma_\Omega$ are $\bold{h_i(t)}=[h_{i,1}, h_{i,2}, h_{i,3} ... h_{i,{T_s\times t}}]$. As an illustrative sample, Figure \ref{100s}(a) shows the $\Omega(t)$ during the corresponding states. Figure \ref{100s}(b) depicts the dynamics and state transitions with $P \in \textbf{T}^{3\times3}$ for $\tau_T=100$s. Note that the initial state is arbitrary and $p(k)=2$ at $t=0$.


\subsection{State Based Power Allocation (SBPA)}
With the knowledge of $P$ by statistical channel state estimation error (s-CSEE) $\sigma_{\textrm{sCSEE}}^2=0$, one way to benefit from CSI for Sat$_1$ is to optimize a resource allocation for $\alpha_k \rightarrow p_k, \forall k$. A state-based power allocation (SBPA) by assigning state priority $\mathcal{P}_k$ for each state duration $\tau_k$ can be given as 
\begin{equation}
    \alpha_k = \frac{\mathcal{P}_k \tau_{t}}{\sum\limits_{k\in K} \mathcal{P}_k \tau_k} ,
\end{equation}
where $\tau_t=\sum\limits_{k\in K} \tau_k$, so that total power consumption has been preserved. Ultimately, transmission power in dBm is $\varrho_k= 10\log(\alpha_k \rho)$. As another fact, achievable data rate ($R$) can be shown as $R_k = \log (1+\frac{\alpha_k \rho \Omega_k}{\sigma^2})$, $\forall k$. 

\section{Numerical Results}
\label{sec4}
This section presents the performance evaluation of the proposed system model through the comparison of two power allocation methods: channel state estimated and state-based power allocated (CSE-SBPA) and the non-state-based conventional method. The simulation parameters used in the numerical results are presented in Table \ref{tab1}. The end-to-end transmission performance of the DSN orbital satellite stations Sat$_1$ and Sat$_2$, which are not affected by atmospheric effects, has been tested. As the study aims to present a generalized model, $\Gamma_{\Omega}$ is arbitrarily selected for each $p(k)$. The generation of $\textbf{T}$ with three steps $p_{jk} \in P$ can be shown as  
\begin{equation}
P=
\begin{bmatrix}
p_{11} & p_{12} & p_{13}\\
p_{21} & p_{22} & p_{23}\\
p_{31} & p_{32} & p_{33}\\
\end{bmatrix}
= 
\begin{bmatrix}
0.8 & 0.1 & 0.1\\
0.5 & 0.3 & 0.2\\
0.7 & 0.25 & 0.05\\
\end{bmatrix}
\end{equation}

Note that state space $S$ has been mapped as $\mathbb{X}:S\rightarrow \mathbb{R}$ and, $\mathbb{X}(\mathrm{BAD})=0$, $\mathbb{X}(\mathrm{MODERATE})=1$ and $\mathbb{X}(\mathrm{GOOD})=2$. 





\begin{figure}[t] 
    \centering
    \includegraphics[width=0.4\textwidth]{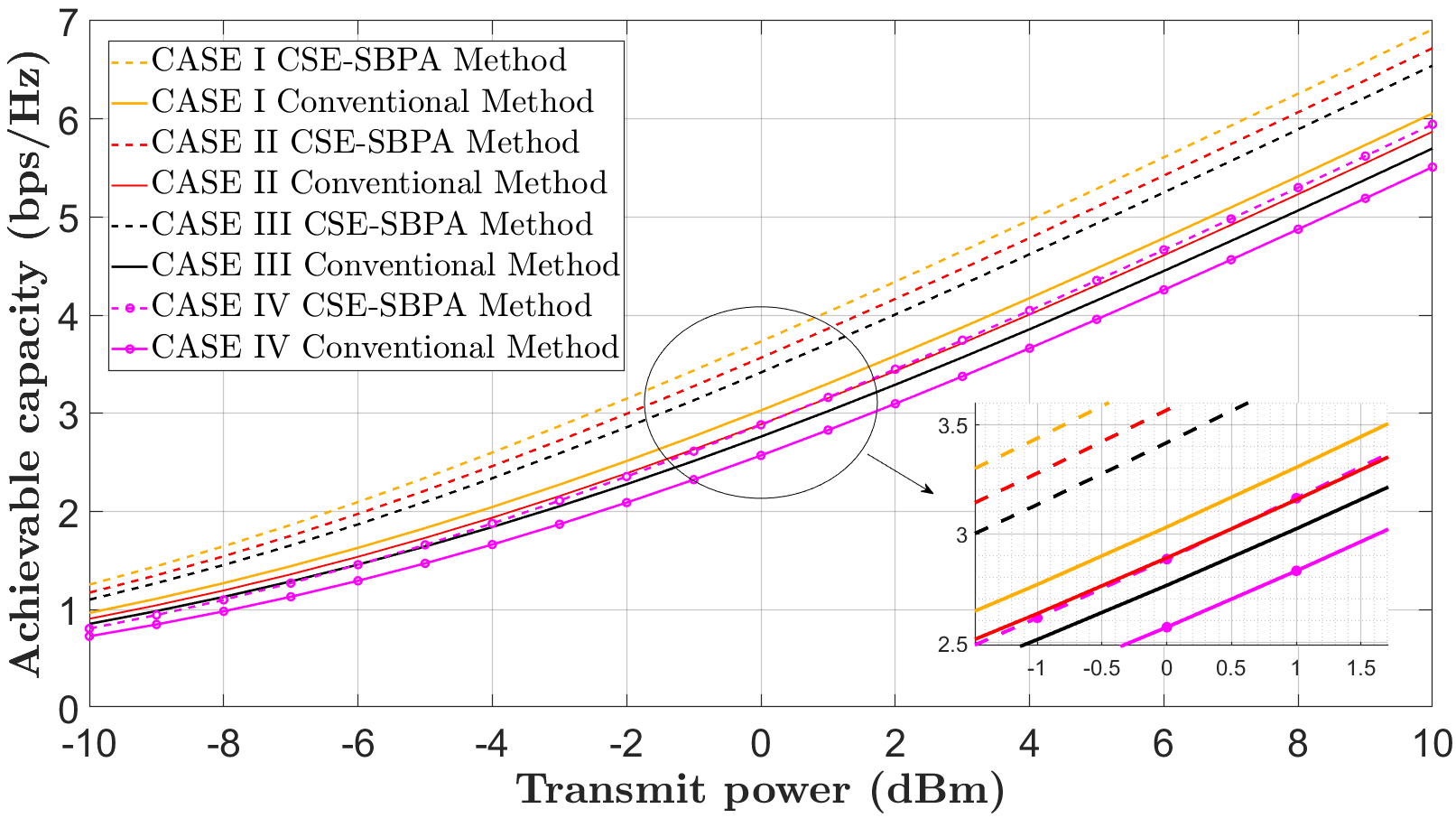}
    \caption{Achievable capacities for time-varying cases.}
    \vspace{-5mm}
    \label{cap}
\end{figure}

\begin{figure}[t] 
    \centering
    \includegraphics[width=0.4\textwidth]{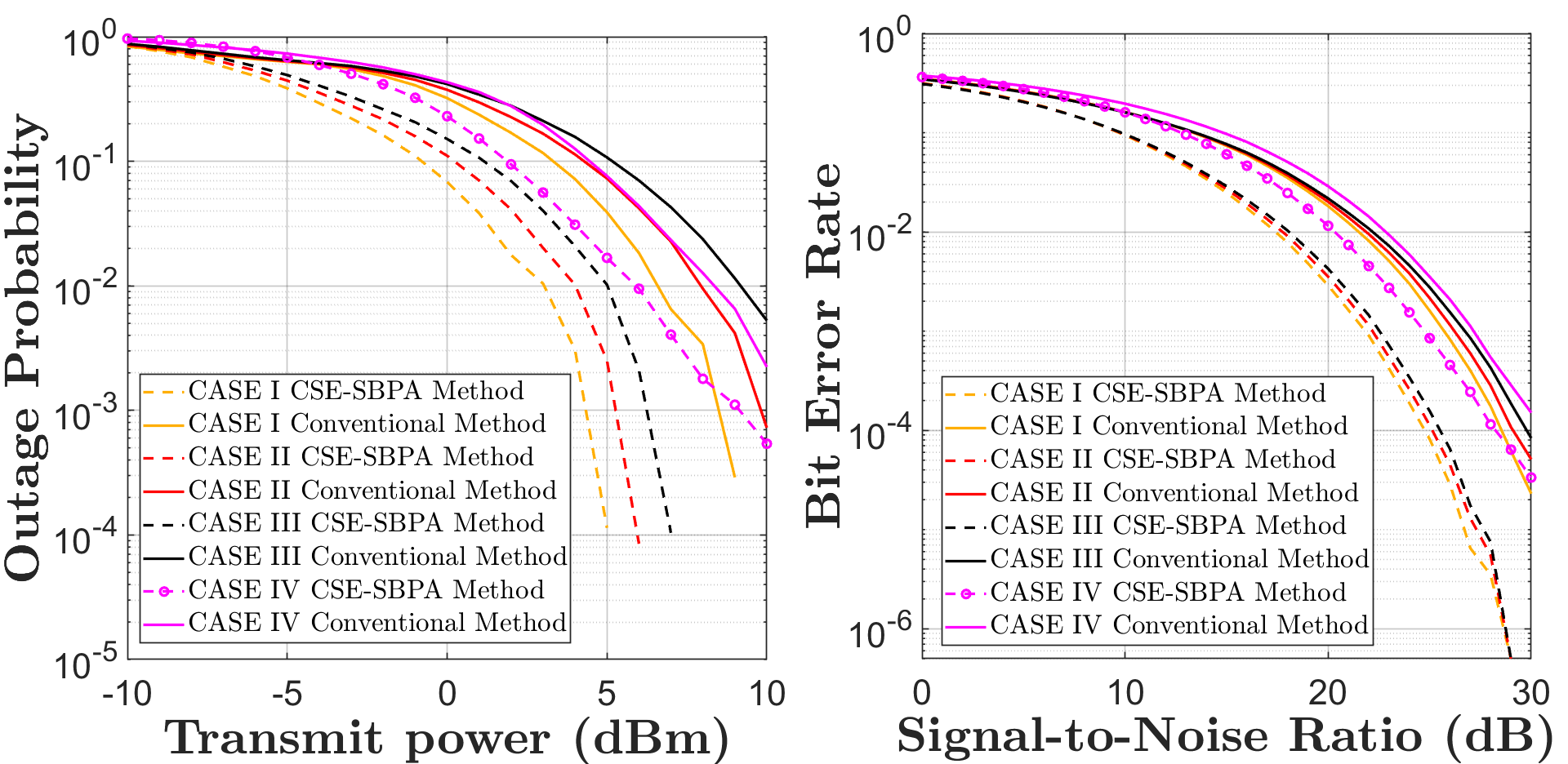}
    \caption{OP and BER for time-varying cases.}
    \vspace{-6mm}
    \label{berop}
\end{figure}

\begin{figure}[t] 
    \centering
\includegraphics[width=0.4\textwidth]{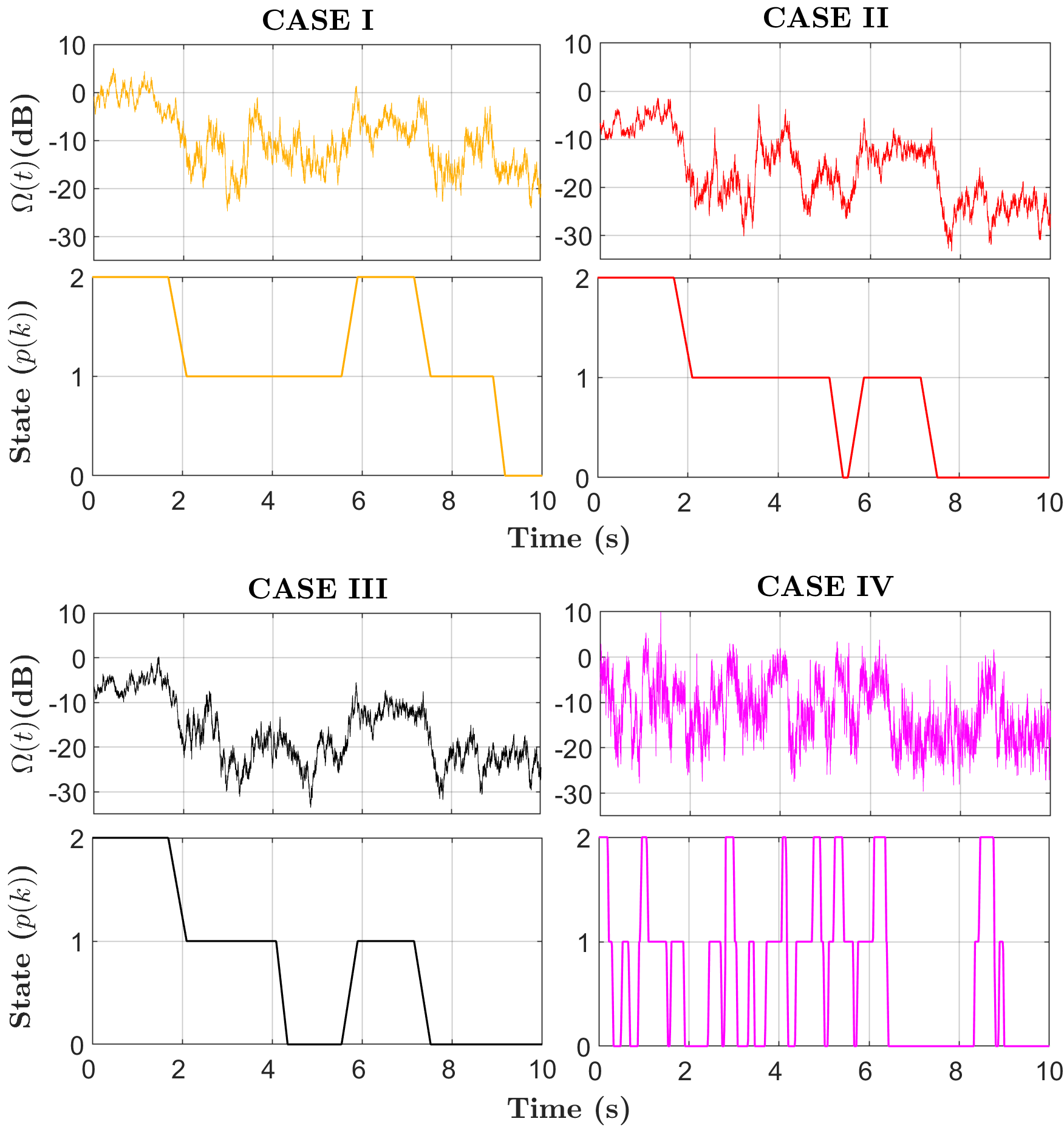}
\vspace{-1mm}
    \caption{Channel state comparison for the defined cases.}
    \vspace{-3mm}
    \label{c1c4}
\end{figure}
To provide a comprehensive analysis, a total of four different cases are considered to evaluate the system performance under various environmental aspects in time-varying simulations:
\begin{description}
\item[\textbf{Case \RNum{1}}:] {~~This case examines the environment where scintillation is tolerably weak where $\Phi$ $>> 3\degree$ with $m\cong 0$, thereby $\gamma \cong \infty$. Yet, usual ISL link challenges still exist.}
\item[\textbf{Case \RNum{2}}:] {~~~This case focuses on investigating the area where the influence of solar activities starts to affect the channel.}
\item[\textbf{Case \RNum{3}}:]{~~~The exact satellite eclipse zone where direct exposure to the sun is inevitable. Note that this is also the point where Sat${_1}$-Sat${_2}$ distance $D_s$ is minimum and equals to $D_1$.}
\item[\textbf{Case \RNum{4}}:] {~~~ In order to examine the relationship between the $\nu_{12}$ and $m$, the velocity difference of \textbf{Case \RNum{1}} was increased and repeated.}
\end{description}

Thereupon, these cases (indexes are denoted with $c$) have been compared with single-state modeled conventional ISL where the receiver completely depends on the i-CSI. In time-varying simulations, the success gap during $p(0)$ state has been recovered by SBPA with $\mathcal{P}^c_{k \in S} = \{2, 1, 0.5\}, \forall c$ and as a result $\alpha^{c}_{k \in S} = \{1.47, 0.73, 0.36\}$ for $c \in \{1,2,3\}$ has been found and used throughout the simulations. It has been observed that $\tau^{c_1}_{k \in S}$ $\cong$ $\tau^{c_2}_{k \in S}$ where $c_1, c_2\in\{1,2,3\}$ with $\tau^c_{k=2}>\tau^{c_1}_{k=1}>\tau^{c_1}_{k=0}$, yet $\tau^{c=4}_{k=1}\cong\tau^{c=4}_{k=2}\cong\tau^{c=4}_{k=3}$ has been measured which implies that $\alpha^{c=4}_{k=1}\cong\alpha^{c=4}_{k=2}\cong\alpha^{c=4}_{k=3}$ which shows that SBPA for \textbf{Case \RNum{4}} was not as efficient as first three cases due to the high variation of $\Omega(t)$ which results with misleading state assumptions.

The channel for each case was structured using a Pseudo Random Sequence (PRS) with a locked seed that has been used to generate fading channel envelope to be able to investigate the same Markovian stochastic process. Figure \ref{cap} shows the $R^c$ for $\exists k$ $\varrho^c_k, \forall c$ obtained by $\exists k$ $\alpha^c_k, \forall c$ with and without the proposed CSE-SBPA method. It is observed that $m$ degrades the $C^c, \forall c$ even though it does not lead to a deep fading. Furthermore, the same figure demonstrates that the proposed method enhances $R^c$, $\forall c$ even when the transmission power is low.    

Lastly, the results of BER and OP analysis for the time-varying scenario with the $R^c_k < \Gamma_k$ outage condition have been presented in Figure \ref{berop}, indicating that CSE-SBPA outperforms the conventional method for all four cases. It has been observed that the overall network performance decreases with an increase in scintillation intensity, despite the decrease in link distance $D_s$. Furthermore, the dominant bottleneck for both OP and BER is still the Doppler frequency $f_{Dop}$. In Figure \ref{c1c4}, all case states along with $\Omega^c(t)$ have been compared, which demonstrates dramatic state transitions during the same $\tau_T$ due to the speed difference.



\section{Conclusion}
Our study focuses on the impact of Doppler shift, solar activities, and thermal noise on ISC models for LoS ISL. We propose a stochastic model that takes into account the angle and time-varying channel states to improve the error, outage, and achievable capacity of such networks. This model leverages statistical knowledge of channel behavior to overcome the unpredictable error bursts caused by these factors.

We demonstrate the effectiveness of our proposed approach by comparing the performance of a $K$-state channel model with that of a single-state ISL system. Our results show that the $K$-state model is superior in terms of error rate, outage probability, and capacity. Additionally, we show that a simple power allocation method, namely the SBPA, is sufficient to improve the performance of the proposed ISL channel model.      

\label{sec5}

\balance
\bibliographystyle{IEEEtran}
\bibliography{reference}
\end{document}